\documentclass[fleqn,usenatbib]{mnras}

\usepackage{newtxtext,newtxmath}

\usepackage[T1]{fontenc}

\DeclareRobustCommand{\VAN}[3]{#2}
\let\VANthebibliography\thebibliography
\def\thebibliography{\DeclareRobustCommand{\VAN}[3]{##3}\VANthebibliography}

\usepackage{graphicx}	
\usepackage{amsmath}	

\title[tribo-ionizing disks]{Ionizing Protoplanetary Disks in Pebble Collisions}

\author[G. Wurm et al.]{
Gerhard Wurm,$^{1}$\thanks{E-mail: gerhard.wurm@uni-due.de (GW)}
Felix Jungmann$^{1}$
and Jens Teiser$^{1}$
\\
$^{1}$University of Duisburg, Faculty of Physics,
Lotharstr. 1,
47057 Duisburg, Germany\\
}

\date{Accepted 2022 July 15. Received 2022 July 11; in original form 2022 April 20}

\pubyear{2022}

\begin{document}
\label{firstpage}
\pagerange{\pageref{firstpage}--\pageref{lastpage}}
\maketitle

\begin{abstract}
We introduce collisions of solids as a new and efficient ionization mechanism for gas in protoplanetary disks, which especially operates in the dense midplane of protoplanetary disks. This idea is sparked by laboratory experiments where we found that charge, which is exchanged by grains in mutual collision (tribocharging), is not tied to their surfaces alone. As kind of collateral effect, charges also become entrained into the gas phase, i.e. collisions ionize the protoplanetary disk. Therefore, solids are not only sinks of charges in disks but also sources.  A first estimate shows that ionization rates in the midplane at 1 AU  {in the range of  $10^{-19} ... 10^{-15} \rm \, s^{-1}$ seem} feasible  {depending on the assumption of rather calm or highly turbulent conditions with radial particle pile up}. 
\end{abstract}




\section{Introduction}

There are well known mechanisms that are regularly invoked to discuss the ionization of gas in protoplanetary disks. First, at sufficiently high temperature, i.e. in the hot inner region  at temperatures beyond 1000 K thermal ionization provides significant amounts of ions \citep{Armitage2011}. Second, high energy radiation of various kind impinging the disk from the outside can be an efficient ionization source. This itself has several components. X-rays generated in the vicinity of the central star are important close to the star and at the disk's surface while ionization by galactic cosmic rays dominates further out and deeper within the disk \citep{Ercolano2013, Glassgold2017, Padovani2018}. And third, radiation by radioactive decay might add ions as well \citep{Umebayashi2009, Cleeves2013, Johansen2018}. 

Many details to describe the charge state in detail are unknown within a complex scheme of ionization, recombination, ion and electron capture by dust and chemistry \citep{Henning2004, Ilgner2006, Schreyer2008, Gammie1996, Dzyurkevich2013, Delage2021, Okuzumi2009,Okuzumi2015,Akimkin2020,Jankovic2021}.

The final ionization states are important for many aspects related to the disk, the gas motion and planet formation, especially for a variety of magnetic interactions within the disk \citep{Balbus1991, Turner2014, Riols2019,Deng2021,Cui2021,Flock2012,Barge2016,Yang2018,Charnoz2021,Ormel2007,Gong2021,Mori2021}.

In the following we will outline a new ionization mechanism and give a simple, first estimate of the ionization rate it might provide in the midplane of protoplanetary disks. All complexities to determine the charge state of the disk mentioned above still remain. We do neither solve nor simplify any of these problems. We "only" add a new way of ionization here which might dominate in parts of a disk at certain times. 
We define the ionization rate $R$ as
\begin{equation}
    R = \frac{1}{N_g} \frac{\Delta N_g}{\Delta t}
    \label{eq.one}
\end{equation}
with the number of gas molecules $\Delta N_g$ which are ionized in a  time interval $\Delta t$ as fraction of the total gas molecule number $N_g$. 

Ionization rates by galactic cosmic rays on the order of $R = 10^{-17} \rm \, s^{-1}$ are reported in the literature \citep{Dalgarno2006, Armitage2010}. Smaller values, e.g. $R < 10^{-19} \rm \, s^{-1}$ are reported for protoplanetary disks in combination with stellar winds reducing the cosmic ray flux \citep{Cleeves2013b, Cleeves2015}. We note, that these values are not discussed in any way further here but only serve as order of magnitude reference to place our new mechanism in context.

\section{Charging disks the other way}

Dust in protoplanetary disks easily grows to sub-mm size in the midplane in sticking collisions
\citep{Wurm1998,Blum2000,Wada2009, Pinilla2013,Testi2014,Misener2019,Wurm2021nat,Hasegawa2021}.
No charge has to be involved in this process yet. Sticking collisions encounter a barrier at this sub-mm stage though. As dust aggregates have become compact by then, they bounce off each other in collisions, known as bouncing barrier \citep{Zsom2010,Guettler2010,Kelling2014, Kruss2017, Demirci2017, Kruss2018}.

\subsection{Detour on collisional charging}

These ideas of particle growth might deflect attention from ionization of the gas phase but just recently, it was suggested that these first barriers in particle growth, i.e. the bouncing barrier might be overcome by collisional charging \citep{Steinpilz2020}. This idea is simple in principle. As grains collide, they charge and these charges can just add another attractive, and in contrast to van der Waals forces, long range force, which can be strong and very efficient to glue dust aggregates together into larger clusters \citep{Marshall2005,Lee2015,Jungmann2018,Matias2018,Haeberle2019,Steinpilz2020, Steinpilz2020a, Xiang2021, Teiser2021, Jungmann&Wurm2021}. 

It is amazing though, that this effect of tribocharging, which is phenomenologically known for a really long time, still escapes a unified explanation. Models range from electron transfer and size dependent charging to exchange of volatiles, i.e. water ions \citep{Duff2008, Lacks2011, Waitukaitis2014,Lee2018,Lacks2019, MendezHarper2021,Kaponig2021}. 
 Size dependent charging and subsequent size separation might trigger lightning and chondrule formation in a heterogenous cloud with large size differences \citep{Desch2000,Muranushi2010,Johansen2018,Spahr2020}.
An important aspect for preplanetary application is though that neither material differences nor size difference are needed for charge transfer. Spherical particles of the same material, which might be considered as identical as they can be, still exchange large amounts of charge \citep{Siu2014, Yoshimatsu2017, Jungmann2018, Steinpilz2020,Jungmann&Wurm2021}. 
Therefore, charging occurs in collisions in any case. In a cloud with similar sized grains, e.g. at the bouncing barrier, the cloud will remain overall neutral and larger cluster can grow \citep{Steinpilz2020, Teiser2021}.

Why we call these thoughts a detour is that this charging does not seem primarily important for the ionization state of the disk's gas. 
So, collisional charging is of emerging importance for grain aggregation, but the connection to disk ionization in terms of molecular ions is not immediately obvious. But are charges generated at the surfaces of grains through tribocharging or collisional charging really restricted to the grains?

\subsection{Collateral charging of gas}

Leakage of charges into the environmental gas during a single collision of solid grains is hardly ever considered.
This might be due to the fact that only little data exist on this.
\citet{Kline2020} were recently the first to publish results on high precision charge measurements in a collision experiment with charge being determined for two particles prior to and after a bouncing collision. The grains were stored in an acoustic trap before their collision and afterwards.  Within the uncertainties, the total charge is conserved in a collision, which they consider as plausibility check of the method. 

It has to be noted though, that, in fact, it is  an important underlying, yet unspoken assumption made in many triboelectric applications, that charge in a collision should be conserved on the solid surfaces. This is quite a reasonable approximation as experience teaches that charge does not easily leave an isolated surface. However, it might only be an approximation and the assumption might break down at least in some situations and as the level of accuracy in the charge balance considered increases as shown below. While a small amount of charge going astray might safely be ignored if the total charge on a grain is the focus, it might be essential for protoplanetary disks, where, in absolute terms, already low ionization rates make a difference. 

\subsubsection{Two particle collision experiments}

We take our recent experiments in \citet{Jungmann2021} where we observe ionization of the gas phase going along with collisional charging  as motivation to estimate the effect for disks.
The experiments by \citet{Jungmann2021} are similar in some sense to \citet{Kline2020}. We also study the charge on two grains before and after a collision. However, we determine the charge by measuring the acceleration of each grain in an external electric field under microgravity conditions \citep{Jungmann2021}. Details on the setup can be found in \citet{Jungmann2018, Steinpilz2020, Jungmann&Wurm2021}.

\begin{figure}
	\includegraphics[width=\columnwidth]{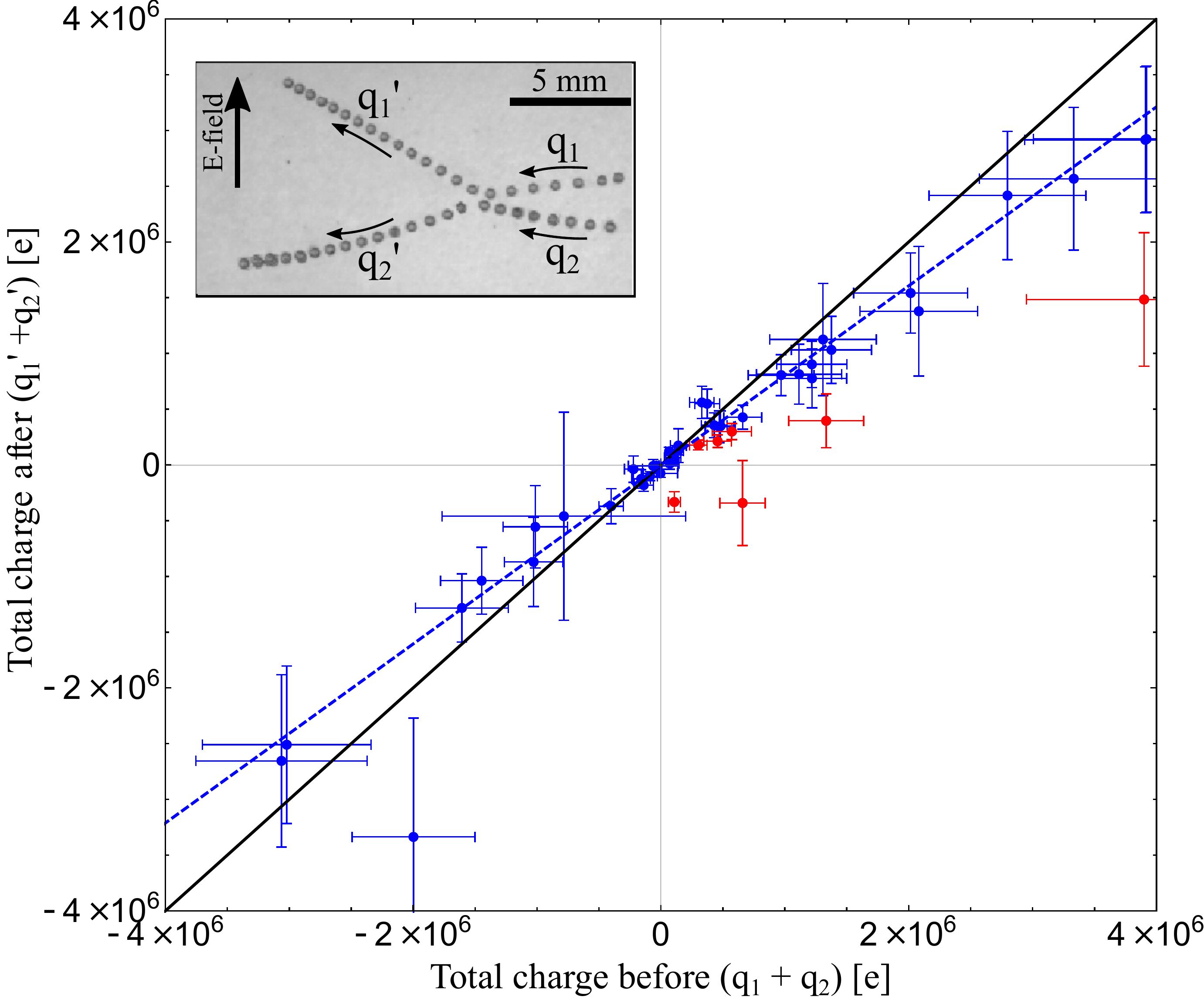}
	\caption{\label{fig.fraction} Total charge on two particles after a collision over charge before the collision. The black line denotes charge conservation. The blue dashed line is a fit to the data. Red data points are not consistent with charge conservation on grains.  {Grains are 217 $\rm \mu m$ in radius.} (from \citet{Jungmann2021}}
\end{figure}

Most of the collisions are, taken individually, also in agreement to charge conservation in a collision within the error bars just like observed by \citet{Kline2020}. However, together the data show a systematic fractional charge loss of about 20\% $\rm \pm 10\%$ independent of net polarity as seen in fig. \ref{fig.fraction}. There are also some collisions which, on their own, are already not in agreement to charge conservation (red data points).

In the given experiments, particles seem to discharge slightly, independent of polarity. A discharge would be plausible as any charge liberated into the environment would be repelled rather than being re-attracted. It also implies though that both polarities of charge are liberated. We note that this does not imply that grains cannot charge in collisions. One grain can still transfer a large amount of any sign from its surface to the other grain, charging the two grains in various ways. It also means that likely even more charge might enter the gas phase, if ions of both polarities are lost to the gas. 

In atmospheric discharges e.g., often space charges are formed, so the amount of ionization might be orders of magnitude higher.
In our experiments, a further fraction of ions of both polarities would not change the net charge balance and will stay undetected in the experiments but would make a great difference for the ionization state in protoplanetary disks. As such, the exact amount would be speculation at this time but we consider our estimates as lower limit for ionization of the gas.

\subsubsection{Mechanisms for pebble collision gas charging}

Even without laboratory experiments, collateral charging of gas in particle collisions seems unavoidable.
There are two aspects that immediately come to mind why gas should charge along with the solids.
The first one is just related to discharge in an atmospheric breakdown.
Grains cannot charge upon collisions forever. A natural limit is reached once any further charge would generate an electric field just large enough for breakdown to occur. This does not have to be a large scale thunderstorm but can occur on much smaller size scale \citep{MendezHarper2016,Matsuyama2018, Wurm2019,MendezHarper2021}. These discharges can be connected to electromagnetic radiation from optical to radio frequencies \citep{Krauss2003,MendezHarper2021, Schoenau2021}.
This is well known on Earth, can be applied to Mars \citep{Kok2009}, and impact ionization in strong electric fields has already been suggested to occur in protoplanetary disks by \citet{Okuzumi2015} to boost ionization of the gas.

The second aspect depends on the nature of the charge carriers. There is growing evidence or at least a number of works suggesting that volatiles like water are responsible for charge transport between grains \citep{Lee2018, Lacks2019, Jungmann2021}. There is water in protoplanetary disks so this will also work in this environment. In any case, it is unlikely that the ions are 100 \% transferred from one grain to the other. Especially in view of pictures where ions are embedded in larger clusters of water molecules, it is rather likely that some charge is diffusing into the environment if water molecules are liberated upon contact.

To highlight another reference for ionization within the Earth atmosphere, it has been measured just recently, that not only the ash particles, but also the gas within volcanic plumes is highly charged \citep{Nicoll2019}, which might also point out the idea of charge generation in the gas phase going along with a dense cloud of solids.
Whatever the reason, there is charge leaking into the gas phase during a collision of solids.

\section{Charging rates in protoplanetary disks}

Our motivation behind this paper is very simple. There is strong experimental evidence of ionization of gas in particle collisions and there is no doubt that there are particle collisions in protoplanetary disks charging grains. So the question is, if we scale these first laboratory results to protoplanetary disks, is this new ionization mechanism of any relevance for the ionization state of protoplanetary disks to motivate further work? 

We give a first estimate 
in a collision scenario needed for the ionization rate as defined in eq. \ref{eq.one}.
To do so, we first give the general equations for the three quantities entering in eq. \ref{eq.one} for a minimum mass nebula and then 
consider possible disk locations and variations of the nebula model depending on the parameters entering in the ionization equation.

There are three quantities that need to be quantified $\Delta t$, $N_g$ and $\Delta N_g$.
Considering all quantities related to a single solid grain, we take $\Delta t$ as the typical time it takes a grain to meet another grain in a collision. This is the collision timescale
\begin{equation}
    \Delta t = \frac{1}{n_p  \sigma  v}.
    \label{eq.dt}
\end{equation}
Here, $n_p$ is the number density of (solid) particles, $\sigma$ is the collision cross section, and $v$ is the collision velocity. 

The number density $n_p$ can be calculated with the absolute dust density $\rho_d$ and the particle mass $m_p$, which is calculated with the bulk density $\rho_p$ assuming spherical agglomerates with the chosen radius $r$ to yield
\begin{equation}
    n_p = \frac{\rho_d}{\frac{4}{3} \pi r^3 \rho_p}
\end{equation}
Note that $\rho_d$ and $\rho_p$ are different particle densities. $\rho_d$ is an average density for the whole volume of the disk's particle cloud, while $\rho_p$ is the bulk density of each particle (agglomerate) in the cloud.

The collision cross section is $\sigma =f \sigma_p$ with the particle geometrical cross section $\sigma_p = \pi r^2$ for spheres. For hard spheres $f$ is 4. 
Due to the long-ranged Coulomb attraction $f$ is typically much larger for charged particles though. Just recently \citet{Jungmann&Wurm2021} measured an enhanced interaction length for charged spheres of 5 on average. For cross sections, this can be translated to $f=25$.

The collision velocity depends on the mechanism considered, which might be radial drift, sedimentation to the midplane or turbulence in our case. We consider radial drift here. In this case,
\begin{equation}
    v = \Delta g \cdot \tau
    \label{vdrift}
\end{equation}
with the gas-grain coupling time $\tau$ and the residual gravity $\Delta g$.  {For high dust to gas ratios this equation needs to be modified } \citep{schneider2019a,schneider2019b,Schneider2021}.  {We do not consider these modifications here.}
This equation further assumes that the gas-grain coupling time is small compared to the orbital timescale, which is the case.
This also assumes as simplification that the drift velocity is also the collision velocity. Certainly, exactly equal grains would not collide but already small variations lead to velocity differences of the same magnitude.
In more detail, the coupling time $\tau$ in the molecular flow regime is given as \citep{Blum1996} 
\begin{equation}
    \tau = \epsilon \frac{m_p}{\sigma_p} \cdot \frac{1}{v_t \rho_g} = \epsilon \frac{ 4 \pi}{3} \cdot \frac{\rho_p}{\rho_g} \cdot \frac{r}{v_t}
\end{equation}
with the particle cross section $\sigma_p$, the gas density $\rho_g$ and the thermal gas velocity $v_t = \sqrt{8 k T / (\pi m_g)}$ with the Boltzmann constant $k = 1.38 \cdot 10^{-23} \, \rm J / K$, gas temperature $T$ and gas molecule mass $m_g$, which is $m_g = 2.34 \rm \, u = 3.88 \cdot 10^{-27} \rm \, kg$.
Furthermore, $\epsilon=0.68$ is an empirical pre-factor found by \citet{Blum1996}. 

The residual gravity in eq. \ref{vdrift} is due to the fact that the gas within the disk is supported by a pressure gradient and is \citep{Weidenschilling1977}
\begin{equation}
    \Delta g = \frac{n k T}{m_g R_s}
\end{equation}
with the distance to the star $R_s$ and the power of the assumed pressure gradient in the disk $n$. 
In total this gives a collision velocity
\begin{equation}
    v =  \epsilon  \frac{4 \pi}{3} \cdot \frac{\rho_p}{\rho_g} \cdot \frac{r}{v_t}  \cdot \frac{n k T}{m_g R_s}
    \label{collvel}
\end{equation}
Putting in numbers from \citet{Weidenschilling1977}, i.e. $\rho_g = 10^{-6} \,\rm kg/m^3$, $T = 600 \,K$ at 1 AU and $n=3$, this consistently becomes $v =2 \,\rm cm/s$ for an $r = 1\, \rm mm$  grain of density $\rho_p = 1000\, \rm kg/m^3$.
A size of 1 mm  is about the size at which particles bounce off each other rather than stick together at cm/s speeds \citep{Kruss2017, Kruss2020}. The bulk density is motivated by assuming silicates with a typical material density of $3000 \rm \, kg/m^3$ and assuming a porosity of the dust aggregates of about 66\% \citep{Meisner2013, Teiser2011}.  So in total, we consider a bulk density of $\rho_p = 1000 \, \rm kg/m^3$ as appropriate.
This gives all quantities to calculate the collision timescale as first parameter in eq. \ref{eq.one}.

As second parameter for eq. \ref{eq.one}, the total number of molecules $N_g$ in the surrounding of each particle is calculated  with the dust to gas ratio $\gamma = \rho_d / \rho_g$ and the ratio of solid particle mass to molecule mass or 

\begin{equation}
    N_g = \frac{m_p}{\gamma m_g} = \frac{4 \pi r^3 \rho_p}{3 \gamma m_g}.
\label{ngallgemein}
\end{equation}

Finally, the number of ions produced $\Delta N_g$ has to be specified.
This is 
\begin{equation}
    \Delta N_g = b q_o r^2.
\label{Nggoeshere}
\end{equation}
The parameter $b$ is supposed to be an efficiency factor including further ion pairs. This is larger than 1 but is currently assumed to be $b = 1$ as most conservative guess in the absence of further data. We include it here nevertheless to highlight that this can change the charge balance strongly. The ion number density released is $q_o = 8 \cdot 10^{14} \, \rm m^{-2}$.  {This is our current estimate from two works by} \citet{Becker2022} and \citet{Jungmann2021}. \citet{Becker2022}  {measured the charge on a 1 mm radius glass sphere for protoplanetary disk pressures to be at least 600 pC or $3.8 \cdot 10^9$ elementary charges.}  We then further assume a constant fraction of the total charge on a grain of $20\%$ to be released \citep{Jungmann2021}. Then  the value of $q_o$  { results if these numbers are put in eq. \ref{Nggoeshere}.} 
Putting all together, we get an ionization rate
\begin{equation}
    R = \frac{3 \epsilon f n \sqrt{k m_g} }{8 \sqrt{2 \pi}\rho_p} \cdot \frac{\gamma^2 \cdot b q_o }{r} \cdot \frac{\sqrt{T}}{R_s}
    \label{master}
\end{equation}

\subsection{Expected ionization rates}

We take eq. \ref{master} as basis to discuss actual values for the ionization rate $R$.
Therefore, this equation is already sorted in 3 terms to see the dependencies and estimate what might be important for the local ionization rate.
The first term only holds constants and quantities which will not vary too much at least not under too extreme conditions. 

The second term holds quantities that might vary throughout the evolution of the disk. First, we place the dust to gas ratio $\gamma$ or its square dependence here. This is a potentially variable quantity, e.g. if an interstellar value of $\gamma = 0.01$ is assumed, ionization rates would be lower by 4 orders of magnitude compared to conditions where the dust to gas ratio is unity. However, aiming at conditions at the bouncing barrier and (sub)-mm particles, these grains have significantly settled and the local dust to gas ration is likely rather close to 1. Certainly, if we considered strong density fluctuations, and if $\gamma = 10$ or more for some time, then locally, the ionization rate can increase by further 4 orders of magnitude or in other term it might fluctuate itself then. It might be noted, that the absolute gas density does not enter. Reducing the gas density would reduce the amount of solids for a constant $\gamma$ but at the same time collision velocities increase at decreasing density compensating a lower particle density at least in the Epstein drag regime. 
 
The second term in eq. \ref{master} also holds the charging fraction in a collision. So far we only have one value but this is a quantity that has to be studied in much more detail in the future.
Particle size is important as well and enters as $1/r$. Our motivation to introduce the mechanism in the first place is that due to the bouncing barrier, particle number densities at one size would be especially high. We expect this to be between $100 \rm \mu m$ and a few mm. Smaller grains just stick and grow. For larger grains, fragmentation becomes important as process. This process and its influence on charge is beyond this study but it is very likely that this provides charge just as efficiently. 

The last term holds the distance and temperature of the disk. If we assume temperature to vary as $R^{-1/2}$ as in the minimum mass nebula by \citet{Hayashi1981} then the total distance dependence goes like $R^{-3/2}$.
So the further inwards, the higher the ionization rate. In the following we will pick three examples and give specific values for $R$. We always use $\gamma = 1$.

\subsubsection{Minimum mass nebula, 1 AU, 280 K}
    These conditions result in $R = 2 \cdot 10^{-19} \rm \, s^{-1}$ for particle sizes of $r = 1 \rm \, mm$. At 1 AU in a minimum mass nebula these are parameter combinations, that might come to mind first. The ionization rate is significant but not too large, i.e. on the level of what short lived radionuclides in an early disk might provide \citep{Umebayashi2009, Johansen2018}. For a dense midplane radionuclides might provide two orders of magnitude more ionization \citep{Umebayashi2009}. However, that depends on the $\rm ^{26}Al$ abundance. E.g. \citet{Malamud2022} discuss rather low amounts of $\rm ^{26}Al$ in comets. Therefore, ionization of the gas by solid collisions might be important whenever radionuclides have been discussed before.
    In view of potentially larger collision velocities from turbulence (see below), the sizes of bouncing grains might rather be smaller. As grain size enters as $1/r$, radial drift would then result in $R = 2 \cdot 10^{-18} \rm \, s^{-1}$ for particle sizes of $r = 0.1 \rm \, mm$. \\
    
\subsubsection{Minimum mass nebula, {0.08 AU}, {1000 K}}
    At this closer location to the star, $R = 4 \cdot 10^{-18} \rm \, s^{-1}$ results for particle sizes of $r = 1 \rm \, mm$. With 1000 K this would connect to an inner region, where thermal ionization might be high enough to trigger the MRI. Again, for smaller grains of $r = 0.1 \rm \, mm$ this would be on order of magnitude more or $R = 4 \cdot 10^{-17} \rm \, s^{-1}$. With more than $10^{-17} \rm \, s^{-1}$ we now find an ionization rate as large as unattenuated cosmic rays would provide. \\
    
\subsubsection{Turbulence, $v = 1 m/s$, 1 AU, $\rho_g = 10^{-6} \, \rm kg m^{-3}$, $r$ = 0.1 mm} 
    These conditions give $R = 10^{-15} \rm \, s^{-1}$, a comparatively high ionization value. The difference is that the above calculations consider the case of rather gentle collision velocities induced by radial drift, while we consider turbulence now. So turbulence strongly changes the picture. We picked one example for turbulent collision velocities only to show the potential. We calculate the ionization rate by still using eq. \ref{master} but now multiplying with the velocity ratio between turbulent and drift collision velocity. Calculated by eq. \ref{collvel} in agreement to the values diplayed in \citet{Weidenschilling1977}, $v_{drift} \sim 1 \rm \, mm/s$ for our case, while the turbulent collision velocities are about 1 m/s for the given grain size of $r = 0.1 \rm \, mm$. These values can e.g. be taken from \citet{Johansen2014} or \citet{Brauer2008} for highly turbulent disks with large $\alpha$ values. 
     {High turbulence implies that dust settling is not very effective in the size range discussed. However, radial drift and particle pile up or streaming instabilities in turbulent disks can still provide large values of $\gamma$} \citep{Johansen2007, Ida2021, Birnstiel2010}. 
    However, even if the solid to gas ratio was smaller, with the high value of $R$ there is room to be still dominating the ionization rate within the midplane.\\
    
\noindent
As summary, if we assume rather conservative calm conditions, ionization rates in the midplane decrease from interstellar cosmic ray values at 0.1 AU to radionuclide values at 1 AU. Further out, its importance might decline but that might depend on the application. For high dust to gas fluctuations, e.g. $\gamma = 10$, ionization rates might locally increase very strongly.
Also, more generally, if collision velocities are set by turbulence, ionization rates might go way up. That does not yet include a potentially higher ionization rate due to ion pairs not seen in the current experiment.

\section{Conclusions}

Laboratory experiments provided clear evidence that collisions between grains always charge them. In fact, this is not a debatable point but rather well known and one route to planetesimal formation \citep{Steinpilz2020, Teiser2021}. As kind of collateral effect though, some charge enters the gas phase. We consider this also a robust finding, even if not yet well constrained quantitatively.

What we propose in view of this evidence is, that this "leakage" of charge into the gas will have significant impact on the ionization state of parts of protoplanetary disks.  Our first estimates indicate that the ionization rate in the midplane of a disk at about 1 AU might range between {  $R \sim 10^{-19}$ and $10^{-15}\rm \, s^{-1}$.} It might be lower in the outer parts but locally, in denser regions it might also be higher and as the experimental charge balance does not count neutral ion pairs, it might also generally be higher. In any case, this value can be much higher compared to the ionization rate by radionuclides or the interstellar ionization rate of cosmic rays.

Therefore, whenever the charge state of the protoplanetary disk (gas) is important, collisions of solids might contribute very significantly. Grains are not only sinks for charges which are already present in the gas phase. Colliding grains, especially at the bouncing barrier, can be very efficient sources for ionized gas in protoplanetary disks. 

\section*{Acknowledgements}

This project is supported by DLR Space Administration with funds provided by the Federal Ministry for Economic Affairs and Climate Action (BMWK) under grant number 50 WM 2142. We also acknowledge the anonymous reviewer whose comments helped to significantly improve the manuscript.

\section*{Data availability}

There are no new data associated with this article.

\bibliography{bibbi}

\begin{thebibliography}{}
\makeatletter
\relax
\def\mn@urlcharsother{\let\do\@makeother \do\$\do\&\do\#\do\^\do\_\do\%\do\~}
\def\mn@doi{\begingroup\mn@urlcharsother \@ifnextchar [ {\mn@doi@}
  {\mn@doi@[]}}
\def\mn@doi@[#1]#2{\def\@tempa{#1}\ifx\@tempa\@empty \href
  {http://dx.doi.org/#2} {doi:#2}\else \href {http://dx.doi.org/#2} {#1}\fi
  \endgroup}
\def\mn@eprint#1#2{\mn@eprint@#1:#2::\@nil}
\def\mn@eprint@arXiv#1{\href {http://arxiv.org/abs/#1} {{\tt arXiv:#1}}}
\def\mn@eprint@dblp#1{\href {http://dblp.uni-trier.de/rec/bibtex/#1.xml}
  {dblp:#1}}
\def\mn@eprint@#1:#2:#3:#4\@nil{\def\@tempa {#1}\def\@tempb {#2}\def\@tempc
  {#3}\ifx \@tempc \@empty \let \@tempc \@tempb \let \@tempb \@tempa \fi \ifx
  \@tempb \@empty \def\@tempb {arXiv}\fi \@ifundefined
  {mn@eprint@\@tempb}{\@tempb:\@tempc}{\expandafter \expandafter \csname
  mn@eprint@\@tempb\endcsname \expandafter{\@tempc}}}

\bibitem[\protect\citeauthoryear{{Akimkin}, {Ivlev}  \& {Caselli}}{{Akimkin}
  et~al.}{2020}]{Akimkin2020}
{Akimkin} V.~V.,  {Ivlev} A.~V.,   {Caselli} P.,  2020, \mn@doi [\apj]
  {10.3847/1538-4357/ab6299}, \href
  {https://ui.adsabs.harvard.edu/abs/2020ApJ...889...64A} {889, 64}

\bibitem[\protect\citeauthoryear{{Armitage}}{{Armitage}}{2010}]{Armitage2010}
{Armitage} P.~J.,  2010, {Astrophysics of Planet Formation}.
Cambridge University Press

\bibitem[\protect\citeauthoryear{{Armitage}}{{Armitage}}{2011}]{Armitage2011}
{Armitage} P.~J.,  2011, \mn@doi [\araa] {10.1146/annurev-astro-081710-102521},
  \href {https://ui.adsabs.harvard.edu/abs/2011ARA&A..49..195A} {49, 195}

\bibitem[\protect\citeauthoryear{{Balbus} \& {Hawley}}{{Balbus} \&
  {Hawley}}{1991}]{Balbus1991}
{Balbus} S.~A.,  {Hawley} J.~F.,  1991, \mn@doi [\apj] {10.1086/170270}, \href
  {https://ui.adsabs.harvard.edu/abs/1991ApJ...376..214B} {376, 214}

\bibitem[\protect\citeauthoryear{{Barge}, {Richard}  \& {Le Diz{\`e}s}}{{Barge}
  et~al.}{2016}]{Barge2016}
{Barge} P.,  {Richard} S.,   {Le Diz{\`e}s} S.,  2016, \mn@doi [\aap]
  {10.1051/0004-6361/201628381}, \href
  {https://ui.adsabs.harvard.edu/abs/2016A&A...592A.136B} {592, A136}

\bibitem[\protect\citeauthoryear{{Becker}, {Steinpilz}, {Teiser}  \&
  {Wurm}}{{Becker} et~al.}{2022}]{Becker2022}
{Becker} T.,  {Steinpilz} T.,  {Teiser} J.,   {Wurm} G.,  2022, \mn@doi
  [\mnras] {10.1093/mnras/stac1320}, \href
  {https://ui.adsabs.harvard.edu/abs/2022MNRAS.tmp.1270B} {}

\bibitem[\protect\citeauthoryear{{Birnstiel}, {Dullemond}  \&
  {Brauer}}{{Birnstiel} et~al.}{2010}]{Birnstiel2010}
{Birnstiel} T.,  {Dullemond} C.~P.,   {Brauer} F.,  2010, \mn@doi [\aap]
  {10.1051/0004-6361/200913731}, \href
  {https://ui.adsabs.harvard.edu/abs/2010A&A...513A..79B} {513, A79}

\bibitem[\protect\citeauthoryear{{Blum}, {Wurm}, {Kempf}  \& {Henning}}{{Blum}
  et~al.}{1996}]{Blum1996}
{Blum} J.,  {Wurm} G.,  {Kempf} S.,   {Henning} T.,  1996, \mn@doi [\icarus]
  {10.1006/icar.1996.0221}, \href
  {https://ui.adsabs.harvard.edu/abs/1996Icar..124..441B} {124, 441}

\bibitem[\protect\citeauthoryear{{Blum} et~al.,}{{Blum}
  et~al.}{2000}]{Blum2000}
{Blum} J.,  et~al., 2000, \mn@doi [\prl] {10.1103/PhysRevLett.85.2426}, \href
  {https://ui.adsabs.harvard.edu/abs/2000PhRvL..85.2426B} {85, 2426}

\bibitem[\protect\citeauthoryear{{Brauer}, {Dullemond}  \& {Henning}}{{Brauer}
  et~al.}{2008}]{Brauer2008}
{Brauer} F.,  {Dullemond} C.~P.,   {Henning} T.,  2008, \mn@doi [\aap]
  {10.1051/0004-6361:20077759}, \href
  {https://ui.adsabs.harvard.edu/abs/2008A&A...480..859B} {480, 859}

\bibitem[\protect\citeauthoryear{{Charnoz}, {Avice}, {Hyodo}, {Pignatale}  \&
  {Chaussidon}}{{Charnoz} et~al.}{2021}]{Charnoz2021}
{Charnoz} S.,  {Avice} G.,  {Hyodo} R.,  {Pignatale} F.~C.,   {Chaussidon} M.,
  2021, arXiv e-prints, \href
  {https://ui.adsabs.harvard.edu/abs/2021arXiv210500456C} {p. arXiv:2105.00456}

\bibitem[\protect\citeauthoryear{{Cleeves}, {Adams}  \& {Bergin}}{{Cleeves}
  et~al.}{2013a}]{Cleeves2013b}
{Cleeves} L.~I.,  {Adams} F.~C.,   {Bergin} E.~A.,  2013a, \mn@doi [\apj]
  {10.1088/0004-637X/772/1/5}, \href
  {https://ui.adsabs.harvard.edu/abs/2013ApJ...772....5C} {772, 5}

\bibitem[\protect\citeauthoryear{{Cleeves}, {Adams}, {Bergin}  \&
  {Visser}}{{Cleeves} et~al.}{2013b}]{Cleeves2013}
{Cleeves} L.~I.,  {Adams} F.~C.,  {Bergin} E.~A.,   {Visser} R.,  2013b,
  \mn@doi [\apj] {10.1088/0004-637X/777/1/28}, \href
  {https://ui.adsabs.harvard.edu/abs/2013ApJ...777...28C} {777, 28}

\bibitem[\protect\citeauthoryear{{Cleeves}, {Bergin}, {Qi}, {Adams}  \&
  {{\"O}berg}}{{Cleeves} et~al.}{2015}]{Cleeves2015}
{Cleeves} L.~I.,  {Bergin} E.~A.,  {Qi} C.,  {Adams} F.~C.,   {{\"O}berg}
  K.~I.,  2015, \mn@doi [\apj] {10.1088/0004-637X/799/2/204}, \href
  {https://ui.adsabs.harvard.edu/abs/2015ApJ...799..204C} {799, 204}

\bibitem[\protect\citeauthoryear{Cui \& Lin}{Cui \& Lin}{2021}]{Cui2021}
Cui C.,  Lin M.-K.,  2021, On the Vertical Shear Instability in Magnetized
  Protoplanetary Disks (\mn@eprint {arXiv} {2105.11151})

\bibitem[\protect\citeauthoryear{{Dalgarno}}{{Dalgarno}}{2006}]{Dalgarno2006}
{Dalgarno} A.,  2006, \mn@doi [Proceedings of the National Academy of Science]
  {10.1073/pnas.0602117103}, \href
  {https://ui.adsabs.harvard.edu/abs/2006PNAS..10312269D} {103, 12269}

\bibitem[\protect\citeauthoryear{{Delage}, {Okuzumi}, {Flock}, {Pinilla}  \&
  {Dzyurkevich}}{{Delage} et~al.}{2021}]{Delage2021}
{Delage} T.~N.,  {Okuzumi} S.,  {Flock} M.,  {Pinilla} P.,   {Dzyurkevich} N.,
  2021, arXiv e-prints, \href
  {https://ui.adsabs.harvard.edu/abs/2021arXiv211005639D} {p. arXiv:2110.05639}

\bibitem[\protect\citeauthoryear{Demirci, Teiser, Steinpilz, Landers, Salamon,
  Wende  \& Wurm}{Demirci et~al.}{2017}]{Demirci2017}
Demirci T.,  Teiser J.,  Steinpilz T.,  Landers J.,  Salamon S.,  Wende H.,
  Wurm G.,  2017, \mn@doi [ApJ] {10.3847/1538-4357/aa816c}, 846, 48

\bibitem[\protect\citeauthoryear{{Deng}, {Mayer}  \& {Helled}}{{Deng}
  et~al.}{2021}]{Deng2021}
{Deng} H.,  {Mayer} L.,   {Helled} R.,  2021, \mn@doi [Nature Astronomy]
  {10.1038/s41550-020-01297-6}, \href
  {https://ui.adsabs.harvard.edu/abs/2021NatAs...5..440D} {5, 440}

\bibitem[\protect\citeauthoryear{{Desch} \& {Cuzzi}}{{Desch} \&
  {Cuzzi}}{2000}]{Desch2000}
{Desch} S.~J.,  {Cuzzi} J.~N.,  2000, \mn@doi [\icarus]
  {10.1006/icar.1999.6245}, \href
  {https://ui.adsabs.harvard.edu/abs/2000Icar..143...87D} {143, 87}

\bibitem[\protect\citeauthoryear{Duff \& Lacks}{Duff \& Lacks}{2008}]{Duff2008}
Duff N.,  Lacks D.~J.,  2008, Journal of Electrostatics, 66, 51

\bibitem[\protect\citeauthoryear{{Dzyurkevich}, {Turner}, {Henning}  \&
  {Kley}}{{Dzyurkevich} et~al.}{2013}]{Dzyurkevich2013}
{Dzyurkevich} N.,  {Turner} N.~J.,  {Henning} T.,   {Kley} W.,  2013, \mn@doi
  [\apj] {10.1088/0004-637X/765/2/114}, \href
  {https://ui.adsabs.harvard.edu/abs/2013ApJ...765..114D} {765, 114}

\bibitem[\protect\citeauthoryear{{Ercolano} \& {Glassgold}}{{Ercolano} \&
  {Glassgold}}{2013}]{Ercolano2013}
{Ercolano} B.,  {Glassgold} A.~E.,  2013, \mn@doi [\mnras]
  {10.1093/mnras/stt1826}, \href
  {https://ui.adsabs.harvard.edu/abs/2013MNRAS.436.3446E} {436, 3446}

\bibitem[\protect\citeauthoryear{{Flock}, {Henning}  \& {Klahr}}{{Flock}
  et~al.}{2012}]{Flock2012}
{Flock} M.,  {Henning} T.,   {Klahr} H.,  2012, \mn@doi [\apj]
  {10.1088/0004-637X/761/2/95}, \href
  {https://ui.adsabs.harvard.edu/abs/2012ApJ...761...95F} {761, 95}

\bibitem[\protect\citeauthoryear{{Gammie}}{{Gammie}}{1996}]{Gammie1996}
{Gammie} C.~F.,  1996, \mn@doi [\apj] {10.1086/176735}, \href
  {https://ui.adsabs.harvard.edu/abs/1996ApJ...457..355G} {457, 355}

\bibitem[\protect\citeauthoryear{{Glassgold}, {Lizano}  \& {Galli}}{{Glassgold}
  et~al.}{2017}]{Glassgold2017}
{Glassgold} A.~E.,  {Lizano} S.,   {Galli} D.,  2017, \mn@doi [\mnras]
  {10.1093/mnras/stx2145}, \href
  {https://ui.adsabs.harvard.edu/abs/2017MNRAS.472.2447G} {472, 2447}

\bibitem[\protect\citeauthoryear{{Gong}, {Ivlev}, {Akimkin}  \&
  {Caselli}}{{Gong} et~al.}{2021}]{Gong2021}
{Gong} M.,  {Ivlev} A.~V.,  {Akimkin} V.,   {Caselli} P.,  2021, arXiv
  e-prints, \href {https://ui.adsabs.harvard.edu/abs/2021arXiv210609525G} {p.
  arXiv:2106.09525}

\bibitem[\protect\citeauthoryear{{G{\"u}ttler}, {Blum}, {Zsom}, {Ormel}  \&
  {Dullemond}}{{G{\"u}ttler} et~al.}{2010}]{Guettler2010}
{G{\"u}ttler} C.,  {Blum} J.,  {Zsom} A.,  {Ormel} C.~W.,   {Dullemond} C.~P.,
  2010, \mn@doi [\aap] {10.1051/0004-6361/200912852}, \href
  {https://ui.adsabs.harvard.edu/abs/2010A&A...513A..56G} {513, A56}

\bibitem[\protect\citeauthoryear{{Haeberle}, {Harju}, {Sperl}  \&
  {Born}}{{Haeberle} et~al.}{2019}]{Haeberle2019}
{Haeberle} J.,  {Harju} J.,  {Sperl} M.,   {Born} P.,  2019, \mn@doi [Soft
  Matter] {10.1039/C9SM01272A}, \href
  {https://ui.adsabs.harvard.edu/abs/2019SMat...15.7179H} {15, 7179}

\bibitem[\protect\citeauthoryear{{Hasegawa}, {Suzuki}, {Tanaka}, {Kobayashi}
  \& {Wada}}{{Hasegawa} et~al.}{2021}]{Hasegawa2021}
{Hasegawa} Y.,  {Suzuki} T.~K.,  {Tanaka} H.,  {Kobayashi} H.,   {Wada} K.,
  2021, \mn@doi [\apj] {10.3847/1538-4357/abf6cf}, \href
  {https://ui.adsabs.harvard.edu/abs/2021ApJ...915...22H} {915, 22}

\bibitem[\protect\citeauthoryear{Hayashi}{Hayashi}{1981}]{Hayashi1981}
Hayashi C.,  1981, \mn@doi [Progress of Theoretical Physics Supplement]
  {10.1143/PTPS.70.35}, 70, 35

\bibitem[\protect\citeauthoryear{{Ida}, {Guillot}, {Hyodo}, {Okuzumi}  \&
  {Youdin}}{{Ida} et~al.}{2021}]{Ida2021}
{Ida} S.,  {Guillot} T.,  {Hyodo} R.,  {Okuzumi} S.,   {Youdin} A.~N.,  2021,
  \mn@doi [\aap] {10.1051/0004-6361/202039705}, \href
  {https://ui.adsabs.harvard.edu/abs/2021A&A...646A..13I} {646, A13}

\bibitem[\protect\citeauthoryear{{Ilgner} \& {Nelson}}{{Ilgner} \&
  {Nelson}}{2006}]{Ilgner2006}
{Ilgner} M.,  {Nelson} R.~P.,  2006, \mn@doi [\aap]
  {10.1051/0004-6361:20053867}, \href
  {https://ui.adsabs.harvard.edu/abs/2006A&A...445..223I} {445, 223}

\bibitem[\protect\citeauthoryear{{Jankovic}, {Owen}, {Mohanty}  \&
  {Tan}}{{Jankovic} et~al.}{2021}]{Jankovic2021}
{Jankovic} M.~R.,  {Owen} J.~E.,  {Mohanty} S.,   {Tan} J.~C.,  2021, \mn@doi
  [\mnras] {10.1093/mnras/stab920}, \href
  {https://ui.adsabs.harvard.edu/abs/2021MNRAS.504..280J} {504, 280}

\bibitem[\protect\citeauthoryear{{Johansen} \& {Okuzumi}}{{Johansen} \&
  {Okuzumi}}{2018}]{Johansen2018}
{Johansen} A.,  {Okuzumi} S.,  2018, \mn@doi [\aap]
  {10.1051/0004-6361/201630047}, \href
  {https://ui.adsabs.harvard.edu/abs/2018A&A...609A..31J} {609, A31}

\bibitem[\protect\citeauthoryear{{Johansen}, {Oishi}, {Mac Low}, {Klahr},
  {Henning}  \& {Youdin}}{{Johansen} et~al.}{2007}]{Johansen2007}
{Johansen} A.,  {Oishi} J.~S.,  {Mac Low} M.-M.,  {Klahr} H.,  {Henning} T.,
  {Youdin} A.,  2007, \mn@doi [\nat] {10.1038/nature06086}, \href
  {https://ui.adsabs.harvard.edu/abs/2007Natur.448.1022J} {448, 1022}

\bibitem[\protect\citeauthoryear{{Johansen}, {Blum}, {Tanaka}, {Ormel},
  {Bizzarro}  \& {Rickman}}{{Johansen} et~al.}{2014}]{Johansen2014}
{Johansen} A.,  {Blum} J.,  {Tanaka} H.,  {Ormel} C.,  {Bizzarro} M.,
  {Rickman} H.,  2014, in {Beuther} H.,  {Klessen} R.~S.,  {Dullemond} C.~P.,
  {Henning} T.,  eds, Protostars and Planets VI. p.~547 (\mn@eprint {arXiv}
  {1402.1344}), \mn@doi{10.2458/azu\_uapress\_9780816531240-ch024}

\bibitem[\protect\citeauthoryear{Jungmann \& Wurm}{Jungmann \&
  Wurm}{2021}]{Jungmann&Wurm2021}
Jungmann F.,  Wurm G.,  2021, \mn@doi [Astronomy and Astrophysics]
  {https://doi.org/10.1051/0004-6361/202039430}

\bibitem[\protect\citeauthoryear{{Jungmann}, {Steinpilz}, {Teiser}  \&
  {Wurm}}{{Jungmann} et~al.}{2018}]{Jungmann2018}
{Jungmann} F.,  {Steinpilz} T.,  {Teiser} J.,   {Wurm} G.,  2018, \mn@doi
  [Journal of Physics Communications] {10.1088/2399-6528/aad0d2}, \href
  {https://ui.adsabs.harvard.edu/abs/2018JPhCo...2i5009J} {2, 095009}

\bibitem[\protect\citeauthoryear{{Jungmann}, {van Unen}, {Teiser}  \&
  {Wurm}}{{Jungmann} et~al.}{2021}]{Jungmann2021}
{Jungmann} F.,  {van Unen} H.,  {Teiser} J.,   {Wurm} G.,  2021, \mn@doi [\pre]
  {10.1103/PhysRevE.104.L022601}, \href
  {https://ui.adsabs.harvard.edu/abs/2021PhRvE.104b2601J} {104, L022601}

\bibitem[\protect\citeauthoryear{Kaponig, M{\"o}lleken, Nienhaus  \&
  M{\"o}ller}{Kaponig et~al.}{2021}]{Kaponig2021}
Kaponig M.,  M{\"o}lleken A.,  Nienhaus H.,   M{\"o}ller R.,  2021, \mn@doi
  [Science Advances] {10.1126/sciadv.abg7595}, 7

\bibitem[\protect\citeauthoryear{{Kelling}, {Wurm}  \& {K{\"o}ster}}{{Kelling}
  et~al.}{2014}]{Kelling2014}
{Kelling} T.,  {Wurm} G.,   {K{\"o}ster} M.,  2014, \mn@doi [\apj]
  {10.1088/0004-637X/783/2/111}, \href
  {https://ui.adsabs.harvard.edu/abs/2014ApJ...783..111K} {783, 111}

\bibitem[\protect\citeauthoryear{{Kline}, {Lim}  \& {Jaeger}}{{Kline}
  et~al.}{2020}]{Kline2020}
{Kline} A.~G.,  {Lim} M.~X.,   {Jaeger} H.~M.,  2020, \mn@doi [Review of
  Scientific Instruments] {10.1063/1.5133049}, \href
  {https://ui.adsabs.harvard.edu/abs/2020RScI...91b3908K} {91, 023908}

\bibitem[\protect\citeauthoryear{{Kok} \& {Renno}}{{Kok} \&
  {Renno}}{2009}]{Kok2009}
{Kok} J.~F.,  {Renno} N.~O.,  2009, \mn@doi [\grl] {10.1029/2008GL036691},
  \href {https://ui.adsabs.harvard.edu/abs/2009GeoRL..36.5202K} {36, L05202}

\bibitem[\protect\citeauthoryear{{Krauss}, {Hor{\'a}nyi}  \&
  {Robertson}}{{Krauss} et~al.}{2003}]{Krauss2003}
{Krauss} C.~E.,  {Hor{\'a}nyi} M.,   {Robertson} S.,  2003, \mn@doi [New
  Journal of Physics] {10.1088/1367-2630/5/1/370}, \href
  {https://ui.adsabs.harvard.edu/abs/2003NJPh....5...70K} {5, 70}

\bibitem[\protect\citeauthoryear{{Kruss} \& {Wurm}}{{Kruss} \&
  {Wurm}}{2018}]{Kruss2018}
{Kruss} M.,  {Wurm} G.,  2018, \mn@doi [\apj] {10.3847/1538-4357/aaec78}, \href
  {https://ui.adsabs.harvard.edu/abs/2018ApJ...869...45K} {869, 45}

\bibitem[\protect\citeauthoryear{{Kruss} \& {Wurm}}{{Kruss} \&
  {Wurm}}{2020}]{Kruss2020}
{Kruss} M.,  {Wurm} G.,  2020, \mn@doi [Planetary Science Journal]
  {10.3847/PSJ/ab93c4}, \href
  {https://ui.adsabs.harvard.edu/abs/2020PSJ.....1...23K} {1, 23}

\bibitem[\protect\citeauthoryear{Kruss, Teiser  \& Wurm}{Kruss
  et~al.}{2017}]{Kruss2017}
Kruss M.,  Teiser J.,   Wurm G.,  2017, \mn@doi [A\&A]
  {10.1051/0004-6361/201630251}, 600, A103

\bibitem[\protect\citeauthoryear{{Lacks} \& {Mohan Sankaran}}{{Lacks} \& {Mohan
  Sankaran}}{2011}]{Lacks2011}
{Lacks} D.~J.,  {Mohan Sankaran} R.,  2011, \mn@doi [Journal of Physics D
  Applied Physics] {10.1088/0022-3727/44/45/453001}, \href
  {https://ui.adsabs.harvard.edu/abs/2011JPhD...44S3001L} {44, 453001}

\bibitem[\protect\citeauthoryear{Lacks \& Shinbrot}{Lacks \&
  Shinbrot}{2019}]{Lacks2019}
Lacks D.~J.,  Shinbrot T.,  2019, \mn@doi [Nature Reviews Chemistry]
  {10.1038/s41570-019-0115-1}, 3, 465

\bibitem[\protect\citeauthoryear{{Lee}, {Waitukaitis}, {Miskin}  \&
  {Jaeger}}{{Lee} et~al.}{2015}]{Lee2015}
{Lee} V.,  {Waitukaitis} S.~R.,  {Miskin} M.~Z.,   {Jaeger} H.~M.,  2015,
  \mn@doi [Nature Physics] {10.1038/nphys3396}, \href
  {https://ui.adsabs.harvard.edu/abs/2015NatPh..11..733L} {11, 733}

\bibitem[\protect\citeauthoryear{{Lee}, {James}, {Waitukaitis}  \&
  {Jaeger}}{{Lee} et~al.}{2018}]{Lee2018}
{Lee} V.,  {James} N.~M.,  {Waitukaitis} S.~R.,   {Jaeger} H.~M.,  2018,
  \mn@doi [Physical Review Materials] {10.1103/PhysRevMaterials.2.035602},
  \href {https://ui.adsabs.harvard.edu/abs/2018PhRvM...2c5602L} {2, 035602}

\bibitem[\protect\citeauthoryear{{Malamud}, {Landeck}, {Bischoff}, {Kreuzig},
  {Perets}, {Gundlach}  \& {Blum}}{{Malamud} et~al.}{2022}]{Malamud2022}
{Malamud} U.,  {Landeck} W.~A.,  {Bischoff} D.,  {Kreuzig} C.,  {Perets} H.~B.,
   {Gundlach} B.,   {Blum} J.,  2022, arXiv e-prints, \href
  {https://ui.adsabs.harvard.edu/abs/2022arXiv220600012M} {p. arXiv:2206.00012}

\bibitem[\protect\citeauthoryear{{Marshall}, {Sauke}  \& {Cuzzi}}{{Marshall}
  et~al.}{2005}]{Marshall2005}
{Marshall} J.~R.,  {Sauke} T.~B.,   {Cuzzi} J.~N.,  2005, \mn@doi [\grl]
  {10.1029/2005GL022567}, \href
  {https://ui.adsabs.harvard.edu/abs/2005GeoRL..3211202M} {32, L11202}

\bibitem[\protect\citeauthoryear{{Matias}, {Shinbrot}  \&
  {Ara{\'u}jo}}{{Matias} et~al.}{2018}]{Matias2018}
{Matias} A.~F.~V.,  {Shinbrot} T.,   {Ara{\'u}jo} N.~A.~M.,  2018, \mn@doi
  [{Physical Rev. E}] {10.1103/PhysRevE.98.062903}, \href
  {https://ui.adsabs.harvard.edu/abs/2018PhRvE..98f2903M} {98, 062903}

\bibitem[\protect\citeauthoryear{{Matsuyama}}{{Matsuyama}}{2018}]{Matsuyama2018}
{Matsuyama} T.,  2018, in The 1st International Conference and Exhibition on
  Powder Technology Indonesia (ICePTi) 2017. p. 020001,
  \mn@doi{10.1063/1.5021189}

\bibitem[\protect\citeauthoryear{Meisner, Wurm, Teiser  \& Schywek}{Meisner
  et~al.}{2013}]{Meisner2013}
Meisner T.,  Wurm G.,  Teiser J.,   Schywek M.,  2013, \mn@doi [A\&A]
  {10.1051/0004-6361/201322083}, 559, A123

\bibitem[\protect\citeauthoryear{{M{\'e}ndez Harper} \& {Dufek}}{{M{\'e}ndez
  Harper} \& {Dufek}}{2016}]{MendezHarper2016}
{M{\'e}ndez Harper} J.,  {Dufek} J.,  2016, \mn@doi [Journal of Geophysical
  Research (Atmospheres)] {10.1002/2015JD024275}, \href
  {https://ui.adsabs.harvard.edu/abs/2016JGRD..121.8209M} {121, 8209}

\bibitem[\protect\citeauthoryear{{M{\'e}ndez Harper}, {Cimarelli}, {Cigala},
  {Kueppers}  \& {Dufek}}{{M{\'e}ndez Harper} et~al.}{2021}]{MendezHarper2021}
{M{\'e}ndez Harper} J.,  {Cimarelli} C.,  {Cigala} V.,  {Kueppers} U.,
  {Dufek} J.,  2021, arXiv e-prints, \href
  {https://ui.adsabs.harvard.edu/abs/2021arXiv210514400M} {p. arXiv:2105.14400}

\bibitem[\protect\citeauthoryear{{Misener}, {Krijt}  \& {Ciesla}}{{Misener}
  et~al.}{2019}]{Misener2019}
{Misener} W.,  {Krijt} S.,   {Ciesla} F.~J.,  2019, \mn@doi [\apj]
  {10.3847/1538-4357/ab4a13}, \href
  {https://ui.adsabs.harvard.edu/abs/2019ApJ...885..118M} {885, 118}

\bibitem[\protect\citeauthoryear{{Mori}, {Okuzumi}, {Kunitomo}  \&
  {Bai}}{{Mori} et~al.}{2021}]{Mori2021}
{Mori} S.,  {Okuzumi} S.,  {Kunitomo} M.,   {Bai} X.-N.,  2021, \mn@doi [\apj]
  {10.3847/1538-4357/ac06a9}, \href
  {https://ui.adsabs.harvard.edu/abs/2021ApJ...916...72M} {916, 72}

\bibitem[\protect\citeauthoryear{{Muranushi}}{{Muranushi}}{2010}]{Muranushi2010}
{Muranushi} T.,  2010, \mn@doi [\mnras] {10.1111/j.1365-2966.2009.15848.x},
  \href {https://ui.adsabs.harvard.edu/abs/2010MNRAS.401.2641M} {401, 2641}

\bibitem[\protect\citeauthoryear{{Nicoll} et~al.,}{{Nicoll}
  et~al.}{2019}]{Nicoll2019}
{Nicoll} K.,  et~al., 2019, \mn@doi [Geophysical Research Letters]
  {10.1029/2019GL082211}, \href
  {https://ui.adsabs.harvard.edu/abs/2019GeoRL..46.3532N} {46, 3532}

\bibitem[\protect\citeauthoryear{{Okuzumi}}{{Okuzumi}}{2009}]{Okuzumi2009}
{Okuzumi} S.,  2009, \mn@doi [\apj] {10.1088/0004-637X/698/2/1122}, \href
  {https://ui.adsabs.harvard.edu/abs/2009ApJ...698.1122O} {698, 1122}

\bibitem[\protect\citeauthoryear{{Okuzumi} \& {Inutsuka}}{{Okuzumi} \&
  {Inutsuka}}{2015}]{Okuzumi2015}
{Okuzumi} S.,  {Inutsuka} S.-i.,  2015, \mn@doi [\apj]
  {10.1088/0004-637X/800/1/47}, \href
  {https://ui.adsabs.harvard.edu/abs/2015ApJ...800...47O} {800, 47}

\bibitem[\protect\citeauthoryear{{Ormel} \& {Cuzzi}}{{Ormel} \&
  {Cuzzi}}{2007}]{Ormel2007}
{Ormel} C.~W.,  {Cuzzi} J.~N.,  2007, \mn@doi [\aap]
  {10.1051/0004-6361:20066899}, \href
  {https://ui.adsabs.harvard.edu/abs/2007A&A...466..413O} {466, 413}

\bibitem[\protect\citeauthoryear{{Padovani}, {Ivlev}, {Galli}  \&
  {Caselli}}{{Padovani} et~al.}{2018}]{Padovani2018}
{Padovani} M.,  {Ivlev} A.~V.,  {Galli} D.,   {Caselli} P.,  2018, \mn@doi
  [\aap] {10.1051/0004-6361/201732202}, \href
  {https://ui.adsabs.harvard.edu/abs/2018A&A...614A.111P} {614, A111}

\bibitem[\protect\citeauthoryear{{Pinilla}, {Birnstiel}, {Benisty}, {Ricci},
  {Natta}, {Dullemond}, {Dominik}  \& {Testi}}{{Pinilla}
  et~al.}{2013}]{Pinilla2013}
{Pinilla} P.,  {Birnstiel} T.,  {Benisty} M.,  {Ricci} L.,  {Natta} A.,
  {Dullemond} C.~P.,  {Dominik} C.,   {Testi} L.,  2013, \mn@doi [\aap]
  {10.1051/0004-6361/201220875}, \href
  {https://ui.adsabs.harvard.edu/abs/2013A&A...554A..95P} {554, A95}

\bibitem[\protect\citeauthoryear{{Riols} \& {Latter}}{{Riols} \&
  {Latter}}{2019}]{Riols2019}
{Riols} A.,  {Latter} H.,  2019, \mn@doi [\mnras] {10.1093/mnras/sty2804},
  \href {https://ui.adsabs.harvard.edu/abs/2019MNRAS.482.3989R} {482, 3989}

\bibitem[\protect\citeauthoryear{Schneider \& Wurm}{Schneider \&
  Wurm}{2019}]{schneider2019b}
Schneider N.,  Wurm G.,  2019, \mn@doi [The Astrophysical Journal Letters]
  {10.3847/2041-8213/ab55e0}, 886, L36

\bibitem[\protect\citeauthoryear{Schneider, Wurm, Teiser, Klahr  \&
  Carpenter}{Schneider et~al.}{2019}]{schneider2019a}
Schneider N.,  Wurm G.,  Teiser J.,  Klahr H.,   Carpenter V.,  2019, \mn@doi
  [The Astrophysical Journal] {10.3847/1538-4357/aafd35}, 872, 3

\bibitem[\protect\citeauthoryear{{Schneider} et~al.,}{{Schneider}
  et~al.}{2021}]{Schneider2021}
{Schneider} N.,  et~al., 2021, \mn@doi [\icarus]
  {10.1016/j.icarus.2021.114307}, \href
  {https://ui.adsabs.harvard.edu/abs/2021Icar..36014307S} {360, 114307}

\bibitem[\protect\citeauthoryear{{Schoenau}, {Steinpilz}, {\textbf{Teiser}}  \&
  {Wurm}}{{Schoenau} et~al.}{2021}]{Schoenau2021}
{Schoenau} L.,  {Steinpilz} T.,  {\textbf{Teiser}} t.,   {Wurm} G.,  2021,
  Gran. Mat., 23, 1

\bibitem[\protect\citeauthoryear{{Schreyer} et~al.,}{{Schreyer}
  et~al.}{2008}]{Schreyer2008}
{Schreyer} K.,  et~al., 2008, \mn@doi [\aap] {10.1051/0004-6361:20079318},
  \href {https://ui.adsabs.harvard.edu/abs/2008A&A...491..821S} {491, 821}

\bibitem[\protect\citeauthoryear{{Siu}, {Cotton}, {Mattson}  \&
  {Shinbrot}}{{Siu} et~al.}{2014}]{Siu2014}
{Siu} T.,  {Cotton} J.,  {Mattson} G.,   {Shinbrot} T.,  2014, \mn@doi [\pre]
  {10.1103/PhysRevE.89.052208}, \href
  {https://ui.adsabs.harvard.edu/abs/2014PhRvE..89e2208S} {89, 052208}

\bibitem[\protect\citeauthoryear{{Spahr} et~al.,}{{Spahr}
  et~al.}{2020}]{Spahr2020}
{Spahr} D.,  et~al., 2020, \mn@doi [\icarus] {10.1016/j.icarus.2020.113898},
  \href {https://ui.adsabs.harvard.edu/abs/2020Icar..35013898S} {350, 113898}

\bibitem[\protect\citeauthoryear{{Steinpilz}, {Joeris}, {Jungmann}, {Wolf},
  {Brendel}, {Teiser}, {Shinbrot}  \& {Wurm}}{{Steinpilz}
  et~al.}{2020a}]{Steinpilz2020}
{Steinpilz} T.,  {Joeris} K.,  {Jungmann} F.,  {Wolf} D.,  {Brendel} L.,
  {Teiser} J.,  {Shinbrot} T.,   {Wurm} G.,  2020a, \mn@doi [Nature Physics]
  {10.1038/s41567-019-0728-9}, \href
  {https://ui.adsabs.harvard.edu/abs/2020NatPh..16..225S} {16, 225}

\bibitem[\protect\citeauthoryear{{Steinpilz}, {Jungmann}, {Joeris}, {Teiser}
  \& {Wurm}}{{Steinpilz} et~al.}{2020b}]{Steinpilz2020a}
{Steinpilz} T.,  {Jungmann} F.,  {Joeris} K.,  {Teiser} J.,   {Wurm} G.,
  2020b, \mn@doi [New Journal of Physics] {10.1088/1367-2630/abae43}, \href
  {https://ui.adsabs.harvard.edu/abs/2020NJPh...22i3025S} {22, 093025}

\bibitem[\protect\citeauthoryear{{Teiser}, {Engelhardt}  \& {Wurm}}{{Teiser}
  et~al.}{2011}]{Teiser2011}
{Teiser} J.,  {Engelhardt} I.,   {Wurm} G.,  2011, \mn@doi [\apj]
  {10.1088/0004-637X/742/1/5}, \href
  {https://ui.adsabs.harvard.edu/abs/2011ApJ...742....5T} {742, 5}

\bibitem[\protect\citeauthoryear{{Teiser}, {Kruss}, {Jungmann}  \&
  {Wurm}}{{Teiser} et~al.}{2021}]{Teiser2021}
{Teiser} J.,  {Kruss} M.,  {Jungmann} F.,   {Wurm} G.,  2021, \mn@doi [\apjl]
  {10.3847/2041-8213/abddc2}, \href
  {https://ui.adsabs.harvard.edu/abs/2021ApJ...908L..22T} {908, L22}

\bibitem[\protect\citeauthoryear{{Testi} et~al.,}{{Testi}
  et~al.}{2014}]{Testi2014}
{Testi} L.,  et~al., 2014, in {Beuther} H.,  {Klessen} R.~S.,  {Dullemond}
  C.~P.,   {Henning} T.,  eds, Protostars and Planets VI. p.~339 (\mn@eprint
  {arXiv} {1402.1354}), \mn@doi{10.2458/azu\_uapress\_9780816531240-ch015}

\bibitem[\protect\citeauthoryear{{Turner}, {Fromang}, {Gammie}, {Klahr},
  {Lesur}, {Wardle}  \& {Bai}}{{Turner} et~al.}{2014}]{Turner2014}
{Turner} N.~J.,  {Fromang} S.,  {Gammie} C.,  {Klahr} H.,  {Lesur} G.,
  {Wardle} M.,   {Bai} X.~N.,  2014, in {Beuther} H.,  {Klessen} R.~S.,
  {Dullemond} C.~P.,   {Henning} T.,  eds, Protostars and Planets VI. p.~411
  (\mn@eprint {arXiv} {1401.7306}),
  \mn@doi{10.2458/azu\_uapress\_9780816531240-ch018}

\bibitem[\protect\citeauthoryear{{Umebayashi} \& {Nakano}}{{Umebayashi} \&
  {Nakano}}{2009}]{Umebayashi2009}
{Umebayashi} T.,  {Nakano} T.,  2009, \mn@doi [\apj]
  {10.1088/0004-637X/690/1/69}, \href
  {https://ui.adsabs.harvard.edu/abs/2009ApJ...690...69U} {690, 69}

\bibitem[\protect\citeauthoryear{{Wada}, {Tanaka}, {Suyama}, {Kimura}  \&
  {Yamamoto}}{{Wada} et~al.}{2009}]{Wada2009}
{Wada} K.,  {Tanaka} H.,  {Suyama} T.,  {Kimura} H.,   {Yamamoto} T.,  2009,
  \mn@doi [\apj] {10.1088/0004-637X/702/2/1490}, \href
  {https://ui.adsabs.harvard.edu/abs/2009ApJ...702.1490W} {702, 1490}

\bibitem[\protect\citeauthoryear{{Waitukaitis}, {Lee}, {Pierson}, {Forman}  \&
  {Jaeger}}{{Waitukaitis} et~al.}{2014}]{Waitukaitis2014}
{Waitukaitis} S.~R.,  {Lee} V.,  {Pierson} J.~M.,  {Forman} S.~L.,   {Jaeger}
  H.~M.,  2014, \mn@doi [\prl] {10.1103/PhysRevLett.112.218001}, \href
  {https://ui.adsabs.harvard.edu/abs/2014PhRvL.112u8001W} {112, 218001}

\bibitem[\protect\citeauthoryear{Weidenschilling}{Weidenschilling}{1977}]{Weidenschilling1977}
Weidenschilling S.~J.,  1977, \mn@doi [MNRAS] {10.1093/mnras/180.2.57}, 180, 57

\bibitem[\protect\citeauthoryear{{Wiebe}, {Semenov}  \& {Henning}}{{Wiebe}
  et~al.}{2004}]{Henning2004}
{Wiebe} D.,  {Semenov} D.,   {Henning} T.,  2004, Baltic Astronomy, \href
  {https://ui.adsabs.harvard.edu/abs/2004BaltA..13..459W} {13, 459}

\bibitem[\protect\citeauthoryear{{Wurm} \& {Blum}}{{Wurm} \&
  {Blum}}{1998}]{Wurm1998}
{Wurm} G.,  {Blum} J.,  1998, \mn@doi [\icarus] {10.1006/icar.1998.5891}, \href
  {https://ui.adsabs.harvard.edu/abs/1998Icar..132..125W} {132, 125}

\bibitem[\protect\citeauthoryear{Wurm \& Teiser}{Wurm \&
  Teiser}{2021}]{Wurm2021nat}
Wurm G.,  Teiser J.,  2021, \mn@doi [Nature Reviews Physics]
  {10.1038/s42254-021-00312-7}, 3, 405

\bibitem[\protect\citeauthoryear{{Wurm}, {Schmidt}, {Steinpilz}, {Boden}  \&
  {Teiser}}{{Wurm} et~al.}{2019}]{Wurm2019}
{Wurm} G.,  {Schmidt} L.,  {Steinpilz} T.,  {Boden} L.,   {Teiser} J.,  2019,
  \mn@doi [\icarus] {10.1016/j.icarus.2019.05.004}, \href
  {https://ui.adsabs.harvard.edu/abs/2019Icar..331..103W} {331, 103}

\bibitem[\protect\citeauthoryear{{Xiang}, {Carballido}, {Matthews}  \&
  {Hyde}}{{Xiang} et~al.}{2021}]{Xiang2021}
{Xiang} C.,  {Carballido} A.,  {Matthews} L.~S.,   {Hyde} T.~W.,  2021, \mn@doi
  [\icarus] {10.1016/j.icarus.2020.114053}, \href
  {https://ui.adsabs.harvard.edu/abs/2021Icar..35414053X} {354, 114053}

\bibitem[\protect\citeauthoryear{{Yang}, {Mac Low}  \& {Johansen}}{{Yang}
  et~al.}{2018}]{Yang2018}
{Yang} C.-C.,  {Mac Low} M.-M.,   {Johansen} A.,  2018, \mn@doi [\apj]
  {10.3847/1538-4357/aae7d4}, \href
  {https://ui.adsabs.harvard.edu/abs/2018ApJ...868...27Y} {868, 27}

\bibitem[\protect\citeauthoryear{{Yoshimatsu}, {Ara{\'u}jo}, {Wurm}, {Herrmann}
   \& {Shinbrot}}{{Yoshimatsu} et~al.}{2017}]{Yoshimatsu2017}
{Yoshimatsu} R.,  {Ara{\'u}jo} N.~A.~M.,  {Wurm} G.,  {Herrmann} H.~J.,
  {Shinbrot} T.,  2017, \mn@doi [Scientific Reports] {10.1038/srep39996}, \href
  {https://ui.adsabs.harvard.edu/abs/2017NatSR...739996Y} {7, 39996}

\bibitem[\protect\citeauthoryear{{Zsom}, {Ormel}, {G{\"u}ttler}, {Blum}  \&
  {Dullemond}}{{Zsom} et~al.}{2010}]{Zsom2010}
{Zsom} A.,  {Ormel} C.~W.,  {G{\"u}ttler} C.,  {Blum} J.,   {Dullemond} C.~P.,
  2010, \mn@doi [\aap] {10.1051/0004-6361/200912976}, \href
  {https://ui.adsabs.harvard.edu/abs/2010A&A...513A..57Z} {513, A57}

\makeatother
\end{thebibliography}
\bibliographystyle{mnras}

\bsp	
\label{lastpage}
\end{document}